\documentclass[12pt]{article}
\usepackage{graphicx}
\usepackage{subfigure}
\begin{document}
\begin{center}
{\Large {\bf A Study of Dynamic Finite Size Scaling Behavior of the Scaling Functions-Calculation of Dynamic Critical Index of Wolff Algorithm }}
\end{center}

\vskip 1.5cm

\centerline{SEMRA G\"UND\"U\c{C}, MEHMET D\.ILAVER ,   
MERAL AYDIN$^{\dagger}$ and Y\.I\u{G}\.IT G\"UND\"U\c{C}}

\centerline{\it Hacettepe University, Physics Department,}
\centerline{\it 06532  Beytepe, Ankara, Turkey }

\vskip 0.5cm

\centerline{\normalsize {\bf Abstract} }

{\small

In this work we have studied the dynamic scaling behavior of two
scaling functions and we have shown that scaling functions obey the
dynamic finite size scaling rules. Dynamic finite size scaling of
scaling functions opens possibilities for a wide range of
applications.  As an application we have calculated the dynamic
critical exponent ($z$) of Wolff's cluster algorithm for $2$-, $3$-
and $4$-dimensional Ising models.  Configurations with vanishing
initial magnetization are chosen in order to avoid complications due
to initial magnetization.  The observed dynamic finite size scaling
behavior during early stages of the Monte Carlo simulation yields $z$
for Wolff's cluster algorithm for $2$-, $3$- and $4$-dimensional Ising
models with vanishing values which are consistent with the values
obtained from the autocorrelations. Especially, the vanishing dynamic
critical exponent we obtained for $d=3$ implies that the Wolff algorithm
is more efficient in eliminating critical slowing down in Monte Carlo
simulations than previously reported.

\vskip 0.5cm

{\small {\it Keywords:} Ising model, scaling functions, dynamic
scaling, time evolution of the magnetization and the scaling
functions, dynamic critical exponent. }

\pagebreak

\section{Introduction}

Finite size scaling and universality arguments have been used to study
the critical parameters of spin systems over two
decades~\cite{Barber83}. Jansen, Schaub and Schmittmann~\cite{Janss89}
showed that for a dynamic relaxation process, in which a system is
evolving according to a dynamics of Model A~\cite{Hoh77} and is
quenched from a very high temperature to the critical temperature, a
universal dynamic scaling behavior within the short-time regime
exists~\cite{Wang97,Zheng98,Zheng00}.  The existence of finite size
scaling even in the early stages of the Monte Carlo simulation has
been tested for various spin
systems~\cite{Zheng98,Zheng00,Jaster99,Luo97,Ying01,
Ozoguz00,Schuelke00,Dilaver03},
the dynamic critical behavior is well-studied and it has been shown
that the dynamic finite size scaling relation holds for the
magnetization and for the moments of the magnetization.  For the
$k^{th}$ moment of the magnetization of a spin system, dynamic finite
size scaling relation can be written as~\cite{Janss89}

\begin{equation}
\label{magnetization}
M^{(k)}(t,\epsilon,{m_0,L})=L^{(-k{\beta}/{\nu})}{\cal{M}}^{(k)}(t/\tau,{\epsilon}L^{1/{\nu}},{m_{0}}L^{x_0})
\end{equation}

where $L$ is the spatial size of the system, $\beta$ and $\nu$ are the
well-known critical exponents, $t$ is the simulation time,
${\epsilon}=(T-{T_c})/{T_c}$ is the reduced temperature and $x_0$ is
an independent exponent which is the anomalous dimension of the
initial magnetization ($m_0$).  In Eq.(\ref{magnetization}) $\tau$ is
the autocorrelation time, $\tau \sim L^z$ and $z$ is the dynamic
critical exponent.

$\;$

The relation given in Eq.(\ref{magnetization}) can be used to study
the known critical exponents as well as exponents $z$ and $x_0$.
Moments of the magnetization have their own anomalous dimensions
($k{\beta}/{\nu}$) And using these quantities (in order to obtain
dynamic exponents $z$ and $x_0$) one may expect some ambiguities due
to correction to scaling and errors on determining the anomalous
dimension of the given thermodynamic quantity. The ambiguities due to
the anomalous dimension of the thermodynamic quantity can be avoided
if one considers quantities which are themselves scaling
functions. Moreover, scaling functions are extremely powerful to
identify the order of the phase transition, as well as locating the
transition point of statistical mechanical systems on finite lattices.

$\;$

In this work we propose that
the dynamic finite size scaling relation also holds for the scaling functions
and the scaling relation can be written similarly to the moments of the
magnetization,

\begin{equation}
\label{Scaling}
O(t,\epsilon,{m_0,L})={\cal{O}}^{(k)}(t/\tau,{\epsilon}L^{1/{\nu}},{m_{0}}L^{x_0}) \;.
\end{equation}

Our aim is to study dynamic finite size scaling behavior of the
scaling functions by using Eq.(\ref{Scaling}).

$\;$

In our calculations two different scaling functions are used.  The
first such quantity is  Binder's cumulant~\cite{Binder81,Binder84,
Binder86}. Binder's cumulant is widely used in order to obtain the
critical parameters as well as to determine the type of the phase
transition. This quantity involves the ratio of the moments of the
magnetization or energy. In this work we have used the definition of
Binder's cumulant which involves the ratio of the moments of the
magnetization. Simplest such quantity can be given as

\begin{equation} 
\label{Binder_Equil}
B_n=\frac {<S^{2 n}>}{<S^n>^2} \;.  
\end{equation} 

In the usual definition of Binder's cumulant, $<S^n>$ is the
thermal average of the $n^{\rm th}$ moment of the configuration
average of the spin. In dynamic case, starting from a totally random
configuration ($m_0=0$), runs are repeated until
predetermined number of iterations are reached for each lattice size.
The thermodynamic quantities are
calculated as the configuration averages at each iteration. In order
to calculate Binder's cumulant iteration by iteration, the
configuration averages of the magnetization and its higher moments
 are calculated.  Various moments of the magnetization
are divided iteration by iteration in order to obtain the required
form of Binder's cumulant,
\begin{equation} 
\label{Binder_Time}
B_n(t)=\frac {<S^{2 n}>(t)}{<S^n>^2(t)} \;.
\end{equation} 
Here the averages are calculated over the configurations obtained
at each  iteration.   
In this work we use Binder's cumulant for $n=2$ by using the 
the relation

\begin{equation} 
\label{Binder_2_Time}
B_2(t)=\frac {<S^{2}>(t)}{<|S|>^2(t)} \;.
\end{equation} 

$\;$

The second quantity is the scaling function ($F$) based on the surface
renormalization. This function is studied in detail for the Ising
model~\cite{oliveira1,oliveira2,oliveira3} and $q$-state Potts
model~\cite{oliveira4,semrascaling,semraprob}. In order to calculate
this function, one considers the direction of the majority of spins of
two parallel surfaces which are $ L/2 $ distance away from each other.
For Ising spins, $F$ can be written in the form~\cite{oliveira3},

\begin{equation}
\label{ScalingFunction_Equil}
 F = < sign[{S}_{i}]sign[{S}_{i+L/2}]>
\end{equation}

where $S_i$ is the sum of the spins in the $i^{\rm th}$ surface.
Similar to the calculations of Binder's cumulant, 
iteration-dependent calculation of $F$ requires 
the configuration averages which 
are obtained for each iteration yielding a Monte Carlo time dependent
expression,

\begin{equation}
\label{ScalingFunction_Time}
 F(t) = < sign[{S}_{i}]sign[{S}_{i+L/2}]>(t) \;.
\end{equation}

$F(t)$ can be used in calculating the dynamic finite size scaling
relation given in Eq.(\ref{Scaling}).

$\;$

In Wolff`s algorithm~\cite{Wolff89a}, only spins belonging to a certain
cluster around the seed spin are considered and updated at each Monte
Carlo step. In equilibrium, the dynamic critical exponent of the
Wolff's algorithm can not be obtained directly from the observed
autocorrelation times (${\tau_W}^{'}$), instead the autocorrelation time
(${\tau_W}$) is governed by the average size ($<C>$ of the clusters.  
$\tau_W$ can be obtained by the relation,

\begin{equation}
\tau_W \;=\; {\tau_W}^{'} \frac{<C>}{L^d}   \;.
\end{equation}

$\;$ The dynamic critical exponents of cluster algorithms are
calculated using the autocorrelation times of spin systems in thermal
equilibrium~\cite{Wolff89a,SwendsenWang87,Ito90,Wolff89b,Heerman90,
Baillie91,Tamayo90}.  In these studies the dynamic critical exponent
is observed to be much less than the value obtained by use of local
algorithms.  For $2$-dimensional Ising model Wolff obtained $z\sim
0.25$~\cite{Wolff89b}.  More recently, Heerman and
Burkitt~\cite{Heerman90} suggested that data are consistent with a
logarithmic divergence, but it is very difficult to distinguish
between the logarithm and a small power~\cite{Baillie91}.  For the
$3$-dimensional case, Tamayo et al~\cite{Tamayo90} calculated the
dynamic critical exponent as $z\sim 0.44(10)$. Wolff calculated a
smaller value of $z=0.28(2)$~\cite{Wolff89b} using energy
autocorrelations.  In $4$-dimensions, Tamayo et al~\cite{Tamayo90}
obtained $z$ with a vanishing value. This result is also consistent
with the mean-field solution for the Ising model in four and higher
dimensions. In a recent publication it has been shown that various
alternative cluster algorithms posses similar dynamic
behavior~\cite{Wang02}

The efficiency of the Wolff`s algorithm is directly related to the
size of the updated clusters, hence the efficiency increases during
the quenching process, as the number of iterations increases.  Both
the average cluster size and susceptibility have the same anomalous
dimension, hence in obtaining $\tau$ from the observed behavior of the
dynamic variable, one can replace $<S^2>$ by $<C>$. In our
calculations both quantities have been used in order to scale time
variable for quantities $B_2(t)$ and $F(t)$ considered.

\section{Simulations and Results}

We have studied dynamic scaling for scaling functions $B_2(t)$ and $F(t)$
for $2$-, $3$- and $4$-dimensional
Ising models evolving in time by using Wolff's algorithm. We have
prepared lattices with vanishing initial magnetization and total random
initial configurations are quenched at the corresponding infinite
lattice critical temperature. We have used the lattices
$L=256,384,512,640$, $L=32,48,64,80$ and $L=16,20,24$ for $2$-, $3$- and
$4$-dimensional Ising models, respectively.  For each lattice size,
independent initial configurations are created. The number of initial
configurations varies depending on
the lattice size. On average, ten bins of one thousand runs, twenty
bins of twenty thousand runs and ten bins of ten thousand runs have
been performed for $2$- $3$-, and $4$- dimensional Ising models,
respectively. Errors are
calculated from the average values for each iteration obtained in
different bins.

$\;$

In the dynamic finite size scaling, for the algorithms in which all
spins are checked for updating the Monte Carlo time, $t$ scales as
$t/{L^z}$. In Wolff's algorithm, one cluster is updated at each
iteration, hence there is a need to use the average number of updated
spins at each iteration. If the time is not scaled by the average
cluster size, using only $L^z$ as a factor shifts the curves towards
each other and curves cross at some point, but scaling can not be
observed. In order to see a good scaling, there is a need to use a
factor which is the average cluster size ($<C>(t)$) or alternatively
$<S^2>(t)$. Dynamic scaling using $<C>(t)$ and $<S^2>(t)$ as the factor
in time scaling results in the same value of the dynamic critical
exponent $z$.  The dynamic critical exponent $z$ is calculated using
the relation
   
\begin{equation}
\label{z}
z=z^{\prime} - (2Y_H-d)  
\end{equation}

which is obtained from the relation

\begin{equation}
\tau=\tau{\prime}<C> 
\end{equation}

where $z^{\prime}$ and $\tau^{\prime}$ are the measured values of the
dynamic critical exponent and the autocorrelation time.  In these
calculations, $Y_H$ is taken as $Y_H={15\over 8}$ (Onsager solution),
$Y_H=2.4808$~\cite{Blote95,Talapov96}, $Y_H=3$ (mean-field solution)
for the $2$-, $3$- and $4$- dimensional models, respectively. Since
$<C>$ and $<S^2>$ scale in the same form, in our presentation we have
scaled time axis with $ t <S^2> / L^{z^{\prime}} $.

$\;$  

In Figure \ref{fig1} we have presented Binder's cumulant $(B_2(t))$
before and after the dynamic finite size scaling for $2$-dimensional
Ising model for the lattice sizes considered. Figure \ref{fig1} a)
shows the time evolution of $B_2(t)$ during the relaxation of the system.
During the relaxation process the correlation length tends to grow 
until it reaches the lattice size. 
When the correlation length approaches the lattice size, Binder's
cumulant exhibits an abrupt change and finally it settles to a new,
long correlation length value. The position of this abrupt change
along time axis depends on the linear size ($L$) of the lattice. But the initial
and the final values are exactly the same for all lattice sizes. In Figure
\ref{fig1} b) the scaling of Binder's cumulant ($B_2(t)$) can be seen.
As it is seen from this figure, $B_2(t)$ scales with time as $t <S^2>(t) /
L^{z^{\prime}}$. As a result of scaling, the value $z^{\prime}=1.725
\pm 0.03$ is obtained from minimizing distances between $B_2(t)$ data for
different lattice sizes.

$\;$

Figure \ref{fig2} shows the surface renormalization function $F(t)$ for
the $2-$dimensional Ising model for the same lattice sizes given in
Figure \ref{fig1}. Figure \ref{fig2} a) shows the simulation data and
Figure \ref{fig2} b) shows the scaling by use of $<S^2>(t)$ as the factor
in time scaling.  This function also exhibits an abrupt change from
initial vanishing value to a certain constant value as the correlation
length reaches the size of the lattice.  As it is seen from this
figure, time to reach the plateau is proportional to the linear size
$(L)$ of the system.  The positions of the abrupt changes for both $F(t)$
and $B_2(t)$ are the same for each lattice size. As in the case of $B_2(t)$
given in Figure \ref{fig1}, a good scaling is observed for the same
value of $z^{\prime}=1.725\pm 0.03$.

Figures \ref{fig3} and \ref{fig4} show the simulation data and the
dynamic scaling for $B_2(t)$ and $F(t)$ for $3$- dimensional Ising model,
respectively. In both figures a) shows the time evolution of the
scaling functions and b) shows the functions after dynamic
scaling. Scaling gives $z^{\prime} = 1.95 \pm 0.05$
for both $B_2(t)$ and $F(t)$ for the $3$-dimensional Ising model. Similarly,
figures \ref{fig5} and \ref{fig6} show the simulation data and the dynamic
scaling for $B_2(t)$ and $F(t)$ for $4$- dimensional Ising model,
respectively. For this model scaling of data for $B_2(t)$ and $F(t)$ results
in $z^{\prime}=2.0 \pm 0.2$ and
$z^{\prime}=2.1 \pm 0.2$, respectively. In all these figures simulation
data for functions $B_2(t)$ and $F(t)$ show the same behavior and scaling is
very good.  The errors in the values of $z^{\prime}$ are obtained from
the largest fluctuations in the simulation data for $B_2(t)$ and
$F(t)$. The values of the dynamic critical exponent $z$ are calculated
using Eq.(\ref{z}) for $2$-, $3$- and $4$- dimensional Ising models
and these values are given in Table 1. The literature values are also
given for comparison.

\section{Conclusion}

Wolff's algorithm is one of the most difficult algorithms to calculate
the dynamic critical exponent. Simply the difficulty arises from the
comparison between the number of updated spins and the total number of
spins. At each iteration only a single cluster is updated. In the
literature, for $2$-, $3$- and $4$-dimensions, small dynamic critical
exponents are
obtained~\cite{Wolff89a,SwendsenWang87,Ito90,Wolff89b,Heerman90,
Baillie91,Tamayo90}, but further studies of the data suggest that for
all three dimensions the dynamic critical exponent of the Ising model
can be considered as zero.  The measurement of the dynamic critical
exponent in thermal equilibrium is extremely difficult, since the
correlation length around the phase transition point is as large as
the size of the lattice. In dynamic finite size scaling, since the
correlation length remains smaller than the lattice size, it is
expected that statistically independent configurations lead to better
statistics since there are no finite size effects.

$\;$

In this work we have considered the dynamic scaling behavior of
Binder's cumulant ($B_2(t)$) and the renormalization function ($F(t)$)
for $2$-, $3$- and $4$- dimensional Ising models. We have observed
that these scaling functions can be used to identify the critical
point and the critical exponents during the initial stages of the
thermalization. In our calculations, we have observed that our results
are consistent with vanishing dynamic critical exponent. Despite the
fact that obtaining good statistics is extremely time consuming for
large lattices, finite size effects do not play any role in obtaining
the results.  One can see from the results of dynamic scaling that
scaling is very good and the errors are very small, hence this method
is a good candidate to calculate the dynamic critical exponent for any
spin model and for any algorithm. The most striking result of our
calculations is that the dynamic critical exponent for $3$-dimensional
Ising model is obtained as $z = 0.02 \pm 0.09 $, instead of previously
reported range of values $z = 0.28-0.44$~\cite{Tamayo90,Wolff89b}. This
is a clear indication that the efficiency of the Wolff's algorithm is
better than previously thought, especially in eliminating critical 
slowing down of Monte Carlo simulations. This means that using this
algorithm, very large lattices at criticality can be considered,
without unusually large statistical errors building up.

\section*{Acknowledgements}

We greatfully acknowledge Hacettepe University Research Fund
\mbox{(Project no : 01 01 602 019)} and Hewlett-Packard's 
Philanthropy Programme.

\pagebreak

\pagebreak

\section*{Table Captions}

Table 1.  

The values of calculated dynamic critical exponents ($z$) (using Eq.(\ref{z}))
for $2$-, $3$- and $4$-dimensional Ising models.  First two coulumns
are the values obtained from scaling functions $B_2(t)$ and $F(t)$,
respectively and the third column includes the literature values.

\section*{Figure Captions}

Figure 1 a) Binder cumulant  data ($B_2(t)$) for $2$-dimensional Ising
Model for linear lattice sizes $L=256$, $384$, $512$, $640$ as a function of
simulation time $t$, b) scaling of  $B_2(t)$ data given in a) using
$<S^2>(t)$ as the factor in time scaling,

$\;$

Figure 2 a) Simulation data for the renormalization function 
($F(t)$) as a function of simulation
time $t$ for $2$-dimensional Ising model for linear lattice sizes
$L=256$, $384$, $512$, $640$, b) scaling of $F(t)$ data given in a) using
$<S^2>(t)$ as a factor in time scaling.

$\;$ 

Figure 3. Simulation data for $B_2(t)$ as a function of simulation
time $t$ for $3$-dimensional Ising model for linear lattice sizes
$L=32$, $48$, $64$, $80$, b) scaling of $B_2(t)$ data given in a) using
$<S^2>(t)$ as the factor in time scaling.

$\;$

Figure 4. Simulation data for $F(t)$ as a function of simulation
time $t$ for $3$-dimensional Ising model for linear lattice sizes
$L=32$, $48$, $64$, $80$, b) scaling of $F(t)$ data given in a) using
$<S^2>(t)$ as the factor in time scaling.

$\;$

Figure 5. Simulation data for $B_2(t)$ as a function of simulation
time $t$ for $4$-dimensional Ising model for linear lattice sizes
$L=16$, $20$, $24$, b) scaling of $B_2(t)$ data given in a) using $<S^2>(t)$
as the factor in time scaling.

$\;$

Figure 6. Simulation data for $F(t)$ as a function of simulation
time $t$ for $4$-dimensional Ising model for linear lattice sizes
$L=16$, $20$, $24$, b) scaling of $F(t)$ data given in a) using $<S^2>(t)$
as the factor in time scaling.

\pagebreak

\begin{center}
\begin{tabular}{|p{2cm}|p{3cm}|p{3cm}|p{4cm}|}
\hline
$~~~~~~d$   &  $z(B_2)       $  & $z(F)$        & $z(Literature)$\\
\hline
$~~~~~~2$   & $ 0.0 \pm 0.05 $  & $ 0.02 \pm 0.05 $ & $ 0-0.4
$ ~~~~~~\cite{Wolff89b,Heerman90,Baillie91}\\
\hline
$~~~~~~3$   &  $0.0 \pm 0.09 $  & $ 0.02 \pm 0.09 $ & $ 0.28-0.44$~\cite{Tamayo90,Wolff89b}
\\
\hline
$~~~~~~4$   & $ 0.0 \pm 0.19  $ & $ -0.13 \pm 0.19$ & $ 0 $~~~~~~~~~~~~~~\cite{Tamayo90}
 \\
\hline
\end{tabular}\\
\vskip 0.5cm
\centerline {Table 1.}  
\end{center}
 
\pagebreak

\begin{figure}
\centering
\subfigure[]{\includegraphics[angle=0,height=4truecm,width=5truecm]{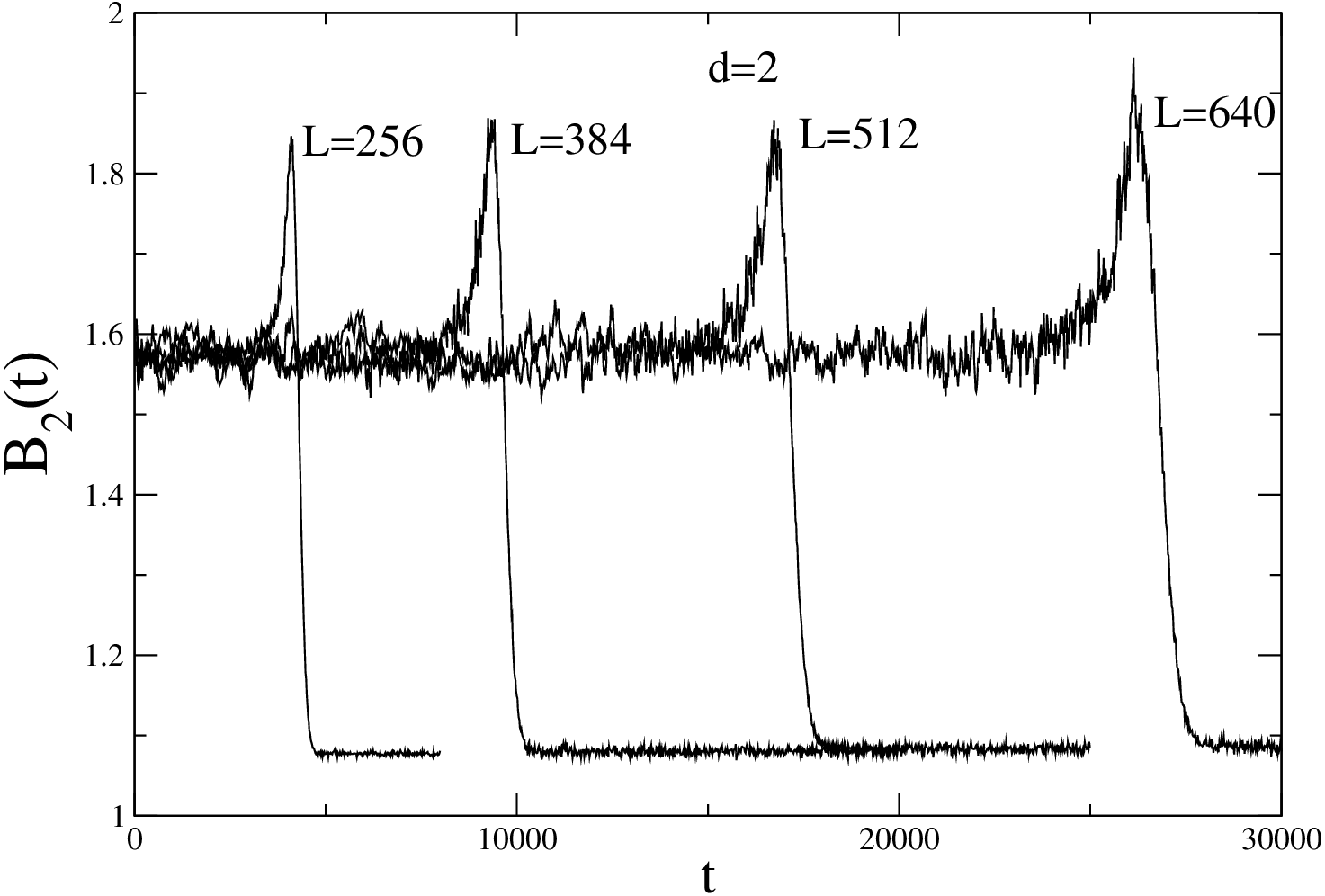}}\\
\subfigure[]{\includegraphics[angle=0,height=4truecm,width=5truecm]{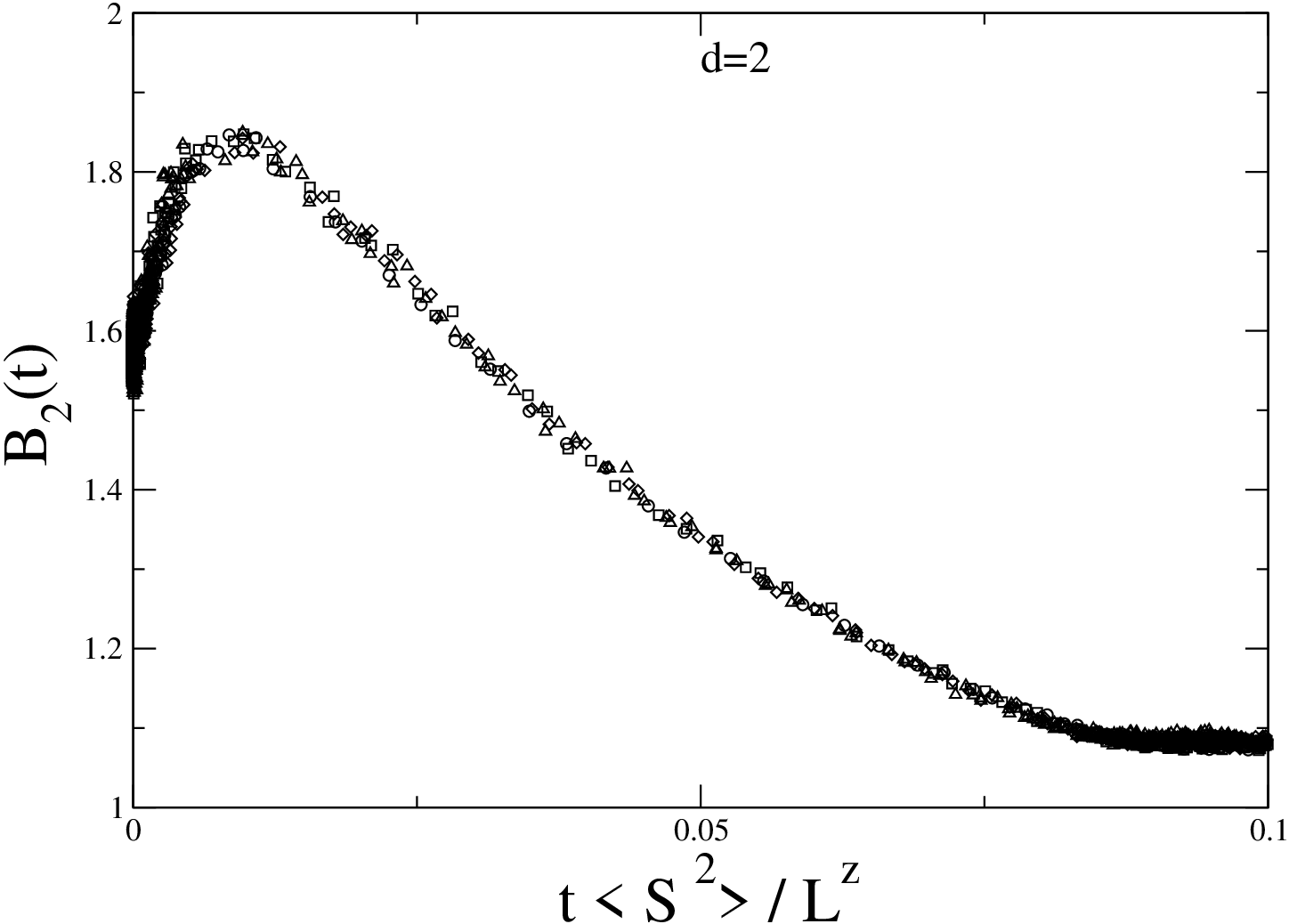}}
\caption{}
\label{fig1}
\end{figure}

\begin{figure}
\centering
\subfigure[]{\includegraphics[angle=0,height=4truecm,width=5truecm]{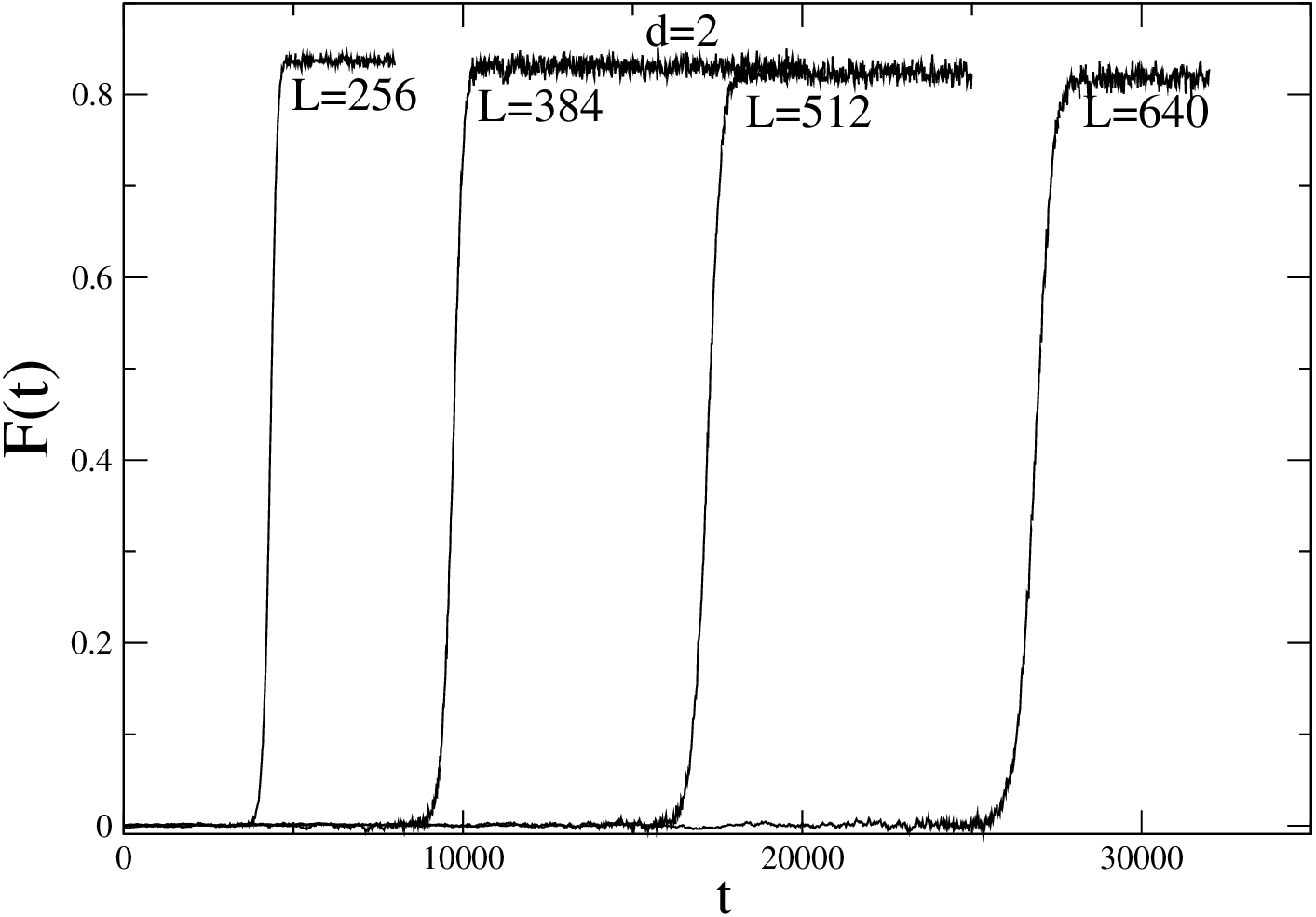}}\\
\subfigure[]{\includegraphics[angle=0,height=4truecm,width=5truecm]{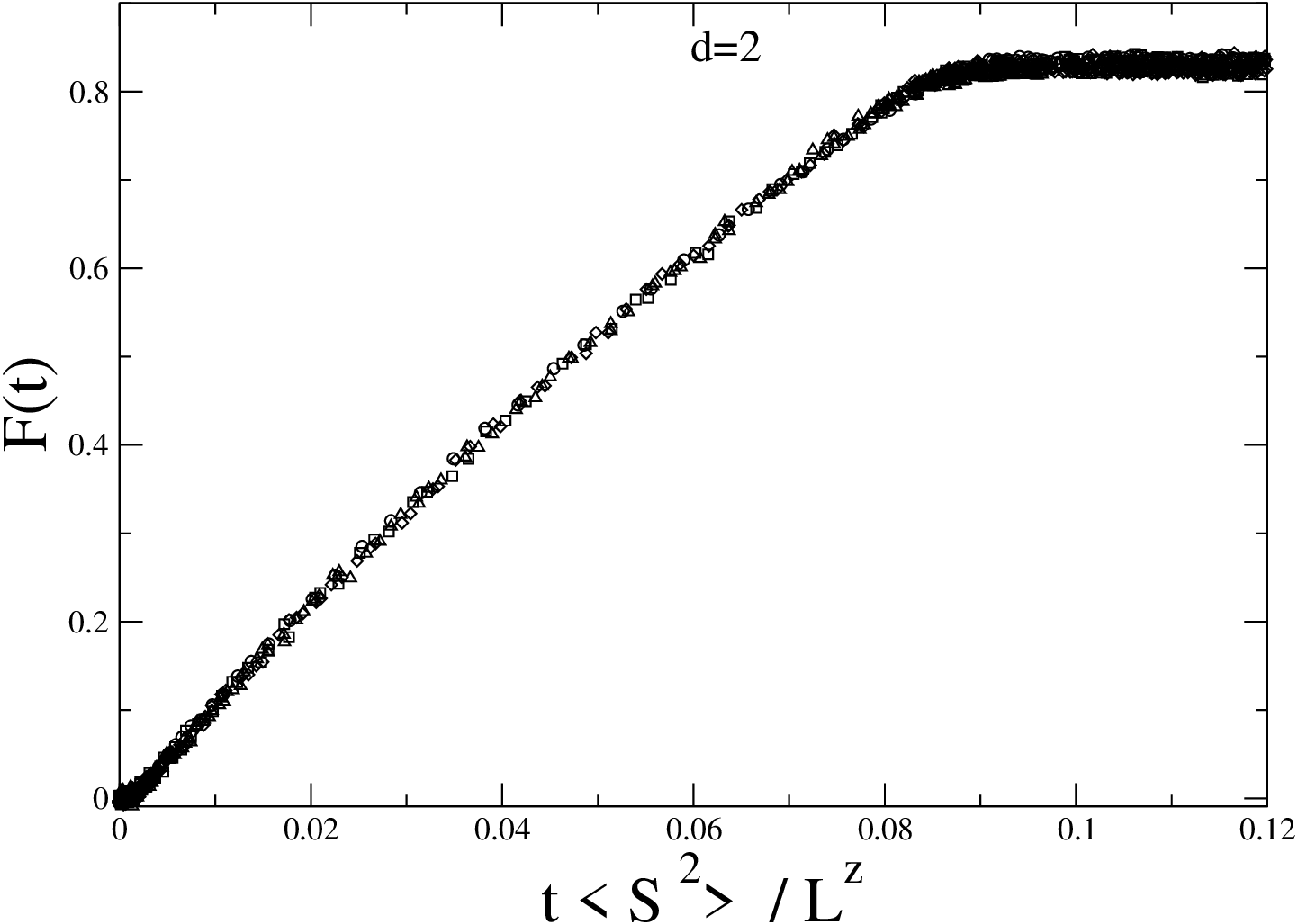}}
\caption{}
\label{fig2}
\end{figure}

\begin{figure}
\centering
\subfigure[]{\includegraphics[angle=0,height=4truecm,width=5truecm]{B2_NS_3d.eps}}\\
\subfigure[]{\includegraphics[angle=0,height=4truecm,width=5truecm]{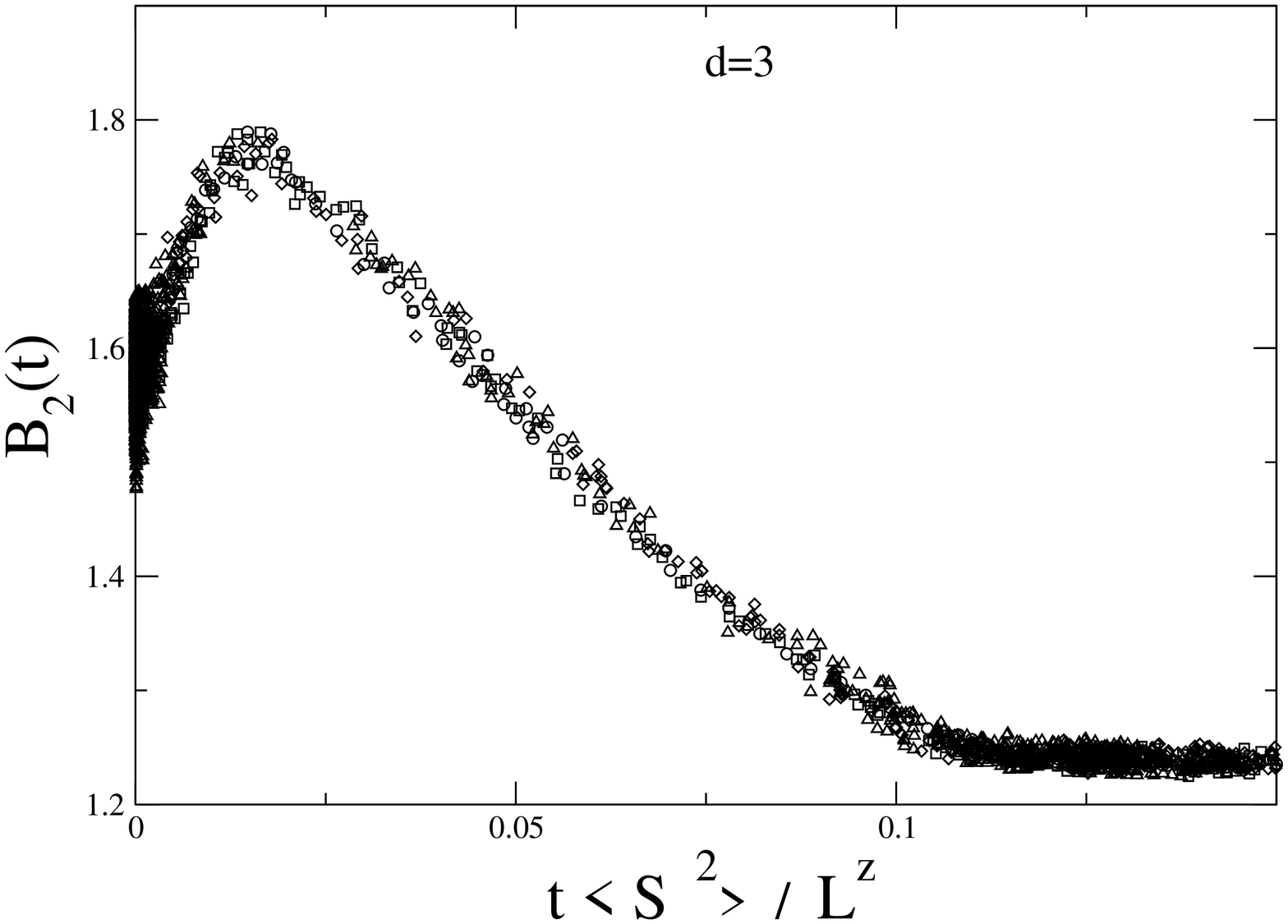}}
\caption{}
\label{fig3}
\end{figure}

\begin{figure}
\centering
\subfigure[]{\includegraphics[angle=0,height=4truecm,width=5truecm]{F_NS_3d.eps}}\\
\subfigure[]{\includegraphics[angle=0,height=4truecm,width=5truecm]{F_WS_3d.eps}}
\caption{}
\label{fig4}
\end{figure}

\begin{figure}
\centering
\subfigure[]{\includegraphics[angle=0,height=4truecm,width=5truecm]{B2_NS_4d.eps}}\\
\subfigure[]{\includegraphics[angle=0,height=4truecm,width=5truecm]{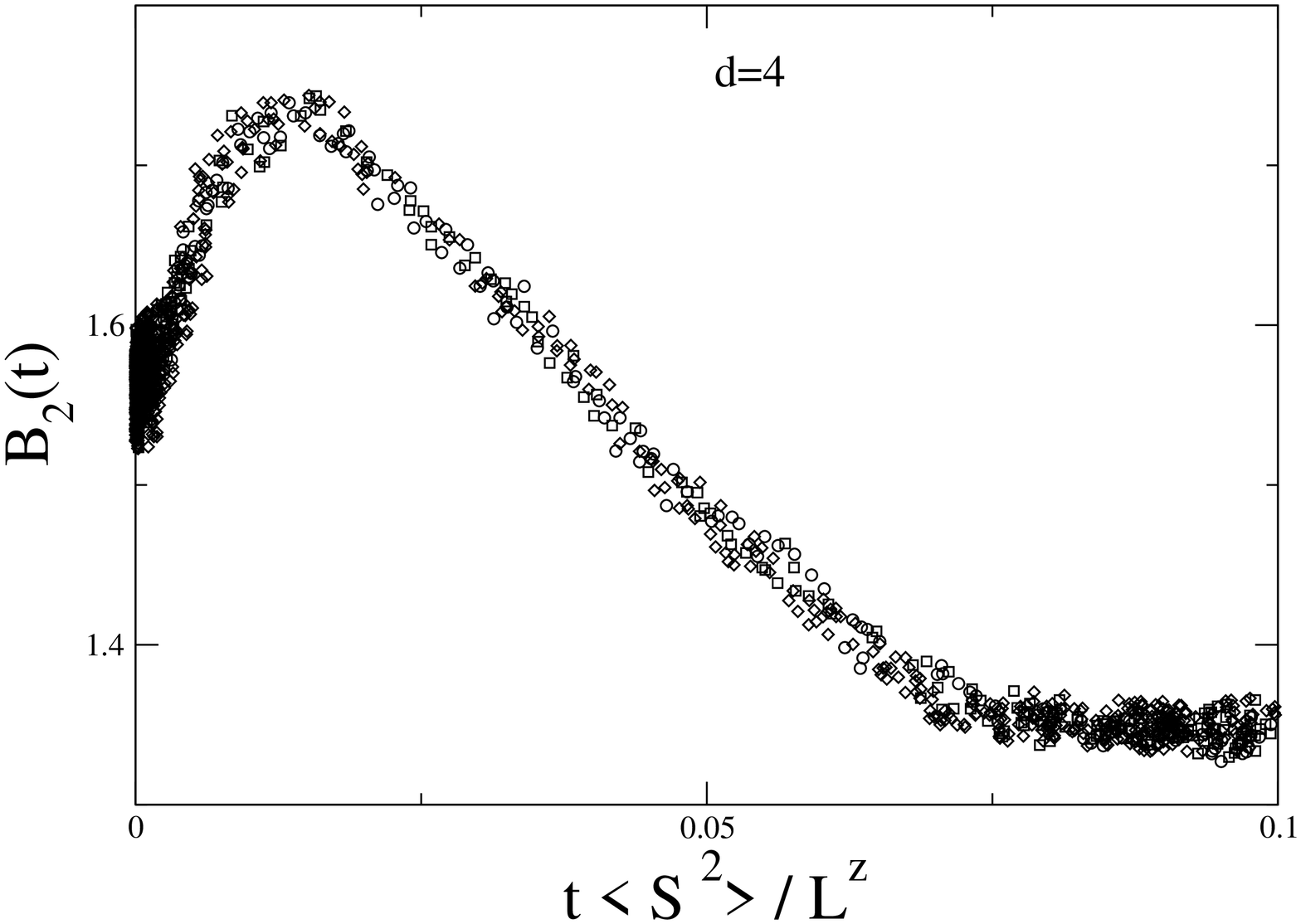}}
\caption{}
\label{fig5}
\end{figure}

\begin{figure}
\centering
\subfigure[]{\includegraphics[angle=0,height=4truecm,width=5truecm]{F_NS_4d.eps}}\\
\subfigure[]{\includegraphics[angle=0,height=4truecm,width=5truecm]{F_WS_4d.eps}}
\caption{}
\label{fig6}
\end{figure}


\begin{thebibliography}{99}

\bibitem{Barber83} M. N. Barber, in: Phase Transition and Critical Phenomena, 
eds. C. Domb and J. L. Lebowitz (Academic, New York, 1983), Vol. 8, p. 146.

\bibitem{Janss89} H. K. Janssen, B. Schaub and B. Schmittmann, Z. Phys. {\bf B73}, 539 (1989).

\bibitem{Hoh77} P. C. Hohenberg and B. I. Halperin, Rev. Mod. Phys. {\bf 49}, 435 (1977).

\bibitem{Wang97} F. G. Wang and C. K. Hu, Phys. Rev. {\bf E56}, 2310 (1997).

\bibitem{Zheng98} B. Zheng, Int. J. Mod. Phys. {\bf B12}, 1419 (1998).

\bibitem{Zheng00} B. Zheng, Physica {\bf A283}, 80 (2000).

\bibitem{Jaster99} A. Jaster, J. Mainville, L. Sch\"ulke and B. Zheng, 
J. Phys. A: Math. Gen. {\bf 32}, 1395 (1999).

\bibitem{Luo97} H. J. Luo and B. Zheng, Mod. Phys. Lett. {\bf B11}, 615 (1997).

\bibitem{Ying01} H. P. Ying, B. Zheng, Y. Yu and S. Trimper, 
Phys. Rev. {\bf E63}, R35101 (2001).

\bibitem{Ozoguz00} B. E. \"{O}zo\u{g}uz, Y. G\"{u}nd\"{u}\c{c} and  M. Ayd{\i}n, Int J. Mod. Phys. {\bf C11}, 553 (2000). 

\bibitem{Schuelke00} L. Sch\"ulke and B. Zheng, Phys. Rev. {\bf E62}, 7482 (2000).

\bibitem{Dilaver03} M. Dilaver, S. G\"und\"u{\c c}, 
M. Ayd{\i}n and Y. G\"und\"u{\c c}, Int. J. Mod. Phys. {\bf C14}, 945 (2003).

\bibitem{Binder81} K. Binder, Phys. Rev. Lett {\bf 47}, 639 (1981).  

\bibitem{Binder84} K. Binder and D. P. Landau,  Phys. Rev. {\bf B30}, 1477 (1984). 

\bibitem{Binder86} M. S. S. Challa, D. P. Landau and  K. Binder, Phys. Rev. {\bf B34}, 1841 (1986).

\bibitem{oliveira1} P. M. C. de Oliveira, Europhys. Lett. {\bf 20}, (1992) 621.

\bibitem{oliveira2} P. M. C. de Oliveira, Physica {\bf A205}, (1994) 101.

\bibitem{oliveira3} J. M. de F. Neto, S. M. de Oliveira, and P. M. C. de Oliveira, Physica {\bf A206}, (1994) 463.

\bibitem{oliveira4} P. M. C. de Oliveira, S. M. de Oliveira, C. E. Cordeiro, and D. Stauffer, J. Stat. Phys. {\bf 80}, (1995) 1433.


\bibitem{semrascaling} S. Demirt\"urk, N. Sefero\u{g}u, M. Ayd{\i}n and Y. G\"und\"u\c{c},  
International Journal of Modern Physics {\bf C12},  (2001) 403.

\bibitem{semraprob} S. Demirt\"urk, Y. G\"und\"u\c{c}, International Journal of Modern Physics {\bf C12}, (2001) 1361.

\bibitem{Wolff89a} U. Wolff, Phys. Rev. Lett. {\bf 62}, 361 (1989).

\bibitem{SwendsenWang87} R. H. Swendsen and J. S. Wang, Phys. Rev. Lett. 
 {\bf 58}, 86 (1987).


\bibitem{Ito90}N. Ito and G. A. Koring, Int. J. Mod. Phys. {\bf C1}, 91 (1990). 

\bibitem{Wolff89b}U. Wolff, Phys. Lett. {\bf B228}, 379 (1989).

\bibitem{Heerman90}D. W. Heerman and A. N. Burkitt, Physica  {\bf A162}, 210 (1990).

\bibitem{Baillie91}C. F. Baillie and P. D. Coddington, Phys. Rev. {\bf B43}, 10617 (1991).

\bibitem{Tamayo90}P. Tamayo, R. C. Brower and W. Klein, J. Stat. Phys. {\bf 58}, 1083 (1990).

\bibitem{Wang02}  J. -S. Wang, O. Kozan and R. H. Swendsen, Phys. Rev. {\bf E66}, 057101 (2002).

\bibitem{Blote95}H. W. J. Bl\"ote, E. Luijten and J. R. Heringa, 
J. Phys. A: Math. Gen. {\bf 28}, 6289 (1995).

\bibitem{Talapov96}A. L. Talapov and H. W. Bl\"ote, J. Phys. A: Math. Gen. {\bf 29}, 5727 (1996).




\end{thebibliography}
\end{document}